\documentclass{article}

   \usepackage[dblblindworkshop, final]{neurips_2025}
 \workshoptitle{Regulatable ML Workshop}

\usepackage[utf8]{inputenc} 
\usepackage[T1]{fontenc}    
\usepackage{hyperref}       
\usepackage{url}            
\usepackage{booktabs}       
\usepackage{amsfonts}       
\usepackage{nicefrac}       
\usepackage{microtype}      
\usepackage{xcolor}         

\title{(When) Should We Delegate AI Governance to AIs? Some Lessons from Administrative Law}

\author{%
  Nicholas Caputo \\
  Oxford Martin AI Governance Initiative\\
  University of Oxford\\
  Oxford OX1 3BD, United Kingdom \\
  \texttt{nick.caputo@oxfordmartin.ox.ac.uk} \\
}

\begin{document}

\maketitle

\begin{abstract}
  Advanced AI systems are now being used in AI governance. Practitioners will likely delegate an increasing number of tasks to them as they improve and governance becomes harder. However, using AI for governance risks serious harms because human practitioners may not be able to understand AI decisions or determine whether they are aligned to the user's interests. Delegation may also undermine governance's legitimacy. This paper begins to develop a principled framework for when to delegate AI governance to AIs and when (and how) to maintain human participation. Administrative law, which governs agencies that are (1) more expert in their domains than the legislatures that create them and the courts that oversee them and (2) potentially misaligned to their original goals, offers useful lessons. Administrative law doctrine provides examples of clear, articulated rules for when delegation can occur, what delegation can consist of, and what processes can keep agencies aligned even as they are empowered to achieve their goals. The lessons of administrative law provide a foundation for how AI governance can use AI in a safe, accountable, and effective way.
\end{abstract}

\section{Introduction}

Frontier AI systems continue to improve in both their capabilities and the scope of tasks they can accomplish \citep{maslej2024artificialintelligenceindexreport, openai2025gpt5}. Because these systems are dual use, these improvements come with risks, especially as AIs become increasingly agentic and self-directed \cite{Bengio_2024}. To keep up with these novel risks, regulation of advanced AI will have to incorporate AI tools. New technologies that create problems have always been met with technical innovations in governance, from the railroad and telegraph enabling both new kinds of torts \cite{atkinson2023telegraph} and new ways of coordinating officials \cite{scott1998seeing} to social media’s harms and the development of algorithmic content moderation \cite{klonick2017new, gillespie2018custodians}.

However, advanced AI presents a special puzzle for governance. Unlike the telegraph or classifier algorithms, these new systems can participate in the design of their own governance institutions because they are capable of the kind of research, analysis, and writing that constitute governance itself. Indeed, everyone working in AI governance, including within frontier companies \cite{stix2025aicloseddoorsprimer}, likely already relies on chatbots at least for tasks like research assistance and editing. If frontier AIs continue to improve, AI-designed governance institutions will likely someday outperform those designed by humans. But, especially given concerns around democratic representation in governance and AI misalignment, simply delegating AI governance decisions to AI without considering the risks of doing so and establishing clear rules for delegation would be dangerous.

We must develop a principled framework for when decisions about how to govern AI should be delegated to AIs. Fortunately, governance has faced a similar problem before. Administrative agencies are expert bodies created by legislatures that delegate power to them to solve specialized problems. They are often empowered to make rules, adjudicate cases, investigate potential violations, and generally act with substantial independence \citep{stewart1975reformation, humphreys1935}. Within their domains, they are more expert than the legislatures that created them and the courts provide judicial review of their actions, creating an “information asymmetry” \citep{huber2006politics, gersen2010designing}. They also sometimes experience “preference divergence” or misalignment from their public purpose, either through capture or internal pressures \cite{Niskanen1971bureaucracy}. Legislatures and courts must therefore find ways to simultaneously empower and oversee agencies while facing a deficit of information and competence compared to the agencies themselves. They seek to resolve this problem through the tools of administrative law.

This paper begins to lay out a framework for delegating governance decisions to AI, drawing on historical lessons from administrative law. It uses five key parts of administrative law doctrine, from nondelegation to "hard look" review, to illustrate how Congress and the courts navigate the problem of delegating authority to expert agencies. In short, it suggests that careful, limited delegation bounded by review processes can help navigate the tradeoff between empowering agencies and risking misalignment. Our analysis is preliminary, sketching commonalities without providing complete guides for delegation, but it suggests that there are useful lessons to draw for this critical governance question.

\section{Background}

Substantial literature exists on whether and when to delegate decisions to AI systems. Researchers have investigated bail and bond determinations in the criminal justice system \citep{angwin2016machine, kleinberg2018human}, child protection hearings \cite{eubanks2018automating}, tenancy determinations \citep{Karpinski20245tenant}, credit provision \citep{oneil2016weapons, hurley2016credit, bartlett2022consumer}, and similar areas. This research has demonstrated the many problems that such delegation can cause, including replicating bias \cite{buolamwini2018gender}, invading privacy \cite{zuboff2019age}, entrenching inequalities \cite{eubanks2018automating}, and causing various dignitary harms \cite{citron2008technological}. However, the existing literature also suggests that in some cases algorithms are better decisionmakers than humans. For example, algorithms may improve on biased or inattentive human judges in bail determinations \cite{kleinberg2018human}, and recent research has shown that AI systems can make better diagnoses than many doctors \cite{esteva2017dermatologist}. Furthermore, algorithms are often (though not always) more explainable and corrigible than humans, so once a problem has been identified in an algorithm, it can be rectified in a way that might not be possible for humans \cite{goel2021accuracy}.

An even vaster administrative law literature covers delegation, deference, and how to create expert governance while preserving accountability and legitimacy. Administrative law begins from the premise that Congress can authorize the creation of administrative agencies empowered to make rules and adjudicate cases in pursuit of specified objectives \cite{strauss1984place, hampton1928}. However, administrative power is  sharply limited by law, which ensures that agencies act in democratic and responsible ways even where they are more expert and capable than Congress and courts \cite{stewart1975reformation}. Some scholars have begun to focus on the use of AI in administration and how the law can help ensure its legitimacy. One key question is whether the use of AI by agencies can meet the requirements of key administrative doctrines including nondelegation, procedural due process, and equal protection \citep{citron2008technological, coglianese2017regulating, solow-niederman2020administering}. Broader analyses consider whether machine learning applications undermine fundamental attributes like the accountability and legitimacy of administration \citep{engstrom2020algorithmic, calo2021automated} and how courts might respond \cite{deeks2019judicial}. 
However, because AI systems that might participate in governance design itself are new, the scholarship has neglected how administrative law could inform AI delegation, focusing instead on how AI might change administrative law.

\section{Administrative Law's Lessons}

Some key lessons emerge from the administrative law literature that might be useful for AI delegation. This list is not exhaustive, but rather illustrative of how inspiration could be drawn.

First, administrative law shows it's possible to navigate tradeoffs between capabilities and alignment similar to those facing AI governance. Administrative agencies must be empowered to perform their governance functions. If Congress or the courts intervene too much in administrative decision-making, especially in areas that are highly technical and specialized, the benefits of delegation will be lost. At the same time, simply letting agencies define and pursue their own objectives without oversight risks them acting contrary to their intended purpose \cite{vermeule2024deference}. The recent debate about \textit{Chevron} deference \cite{chevron1984} concerned this question, and the Supreme Court decided that deferring to agency interpretations of their own authorizing statutes gave agencies too much autonomy with too little oversight \cite{loperbright2024}. As a field, administrative law aims to fine-tune the balance between capabilities and alignment of agencies as they act. AI governance already faces similar problems of "information asymmetry" and "preference divergence" to administrative law. It currently lacks, however, the tools to build oversight that would allow us to pick the optimal spot on the tradeoff curve, as administrative law aims to do.

Second, delegation should be done for specific and limited purposes, and certain areas might be best marked out as non-delegable. Congress' power to delegate authority to administrative agencies is limited. It must provide an "intelligible principle" that guides the agency's action rather than simply providing a grant of authority \cite{hampton1928}. Furthermore, for so-called "major questions" of significant importance, Congress must be extremely explicit about its delegation \citep{fdavbrownwilliamson2000, westvirginiavepa2022}. Under the Constitution, delegation should be a limited tool that supplements Congress' ability to perform its functions rather than a full grant of power to an external body \cite{strauss1984place}. Governance delegations to AIs might need to be similarly limited in scope and context. Existing AI laws like the EU AI Act already distinguish the kinds of functions AI systems can perform \cite{euai_act2024}. Extending these concepts to governance, we might decide that key mechanisms like critical risk thresholds or the use of autonomous AI in the military must be designed by humans or by AIs with significant human instruction and oversight because of their significance for human life.

Third, certain process requirements can help compensate for a lack of substantive understanding of why an agent is deciding in a particular way. Because administrative agencies are expert bodies operating in specialized fields, other branches of government struggle to provide oversight \cite{landis1938administrative}. Indeed, Congress' inability to understand what is happening in a given domain is often a key reason it creates an administrative agency in the first place \cite{epstein1999delegating}. Courts providing judicial review similarly find it difficult to provide substantive review of agency decisions that they may lack the context to understand \cite{chevron1984}. 
But this lack of understanding does not mean that oversight is impossible. Administrative law uses procedural requirements placed on agencies to make up for the lack of substantive knowledge. For example, courts can check whether agencies met the requirements that they give reasons for their actions and consider the whole record of evidence when deciding, even if they cannot judge the correctness of the ultimate decisions \cite{statefarm1983}. Tools like "hard look review" provide sufficient oversight of administrative agencies to keep them accountable and ensure the legitimacy of their actions. Agencies are required to give the true reasons why they took their actions, not just make up plausible justifications after the fact \cite{chenery1947}. Requiring that AI systems similarly provide reasons for their decisions and follow established decision-making processes would help people understand the governance decisions they make or at least provide a partial guarantee that the choices were made in a pro-social way. Interpretability, chain-of-thought monitoring, and similar techniques \citep{rudin2019stop, wei2022chain, korbak2025chainthoughtmonitorabilitynew} could even provide a more complete guarantee of legitimacy in the AI context than we get in the context of human administrative agencies.

Fourth, public participation in decision-making enhances both its quality and its legitimacy. Administrative agencies undergo significant public process when making new rules and adjudicating cases. The "notice and comment" process ensures that concerned parties are notified and given the chance to participate in shaping or stopping actions that might affect them \cite{apa1946, novascotia1977}. Adjudications must meet various procedural due process requirements that seek to guarantee that no rights are violated in deciding the case \cite{mathews1976}. Furthermore, public participation provides useful new inputs to decision-making that improve how agencies carry out their work. Those affected by an agency action likely have special local knowledge that would be useful for the agency to consider when making its decision \cite{stewart1975reformation}. Including public voices provides the agencies with new information that might change what they think is the best course of action.
Delegation of AI governance should also include public participation in the rule-making process. AIs making decisions about governance should be required to take human input into account and address it in comprehensive, rational, and rule-bound ways. Such participation would ensure some degree of legitimacy in the process \cite{wilftownsend2025generated} and also ensure that human voices and perspectives shaped how AI is governed.

Finally, administrative law teaches that it is not always a bad idea to rely on the comparative expertise and competence of others. Congress would have to entirely reconstruct itself to serve the variety of functions that the administrative state currently performs, and in doing so, it would undermine its own purpose as a representative legislature. The complex problems of modern life require sophisticated forms of governance that come with tradeoffs, but administrative law and similar institutions can help us make better decisions about how to handle those tradeoffs \cite{landis1938administrative}. AI will have to be a part of the solution to AI's problems. The challenges of AI will require new forms of government just as the new problems created by technological progress in the early twentieth century demanded the creation of the administrative state \cite{mccraw1984prophets}. In fact, AI might enable better governance than is possible with human-run agencies. The capabilities/alignment tradeoff that governance of administrative agencies faces has been relatively immutable over time, a product of human psychology and small advances in technology. In AI, however, advances in the fields of capabilities and alignment, giving people greater insight into why AI systems act than is possible with humans, may create the opportunity for more effective and responsible government. Combining the technical promise of AI with the insights of administrative law could open up powerful new opportunities for governance.

\section{Example Applications}

First, AI risk managers must set thresholds for when to conduct comprehensive evaluations of the risks presented by their systems \citep{euai_act2024, OpenAI-Preparedness-Framework-v2-2025, Anthropic-RSP-v2_2-2025}. Setting thresholds requires understanding system capabilities, the external risk environment, and how different risks can emerge across a variety of domains from biology to cybersecurity \cite{anderljung2023frontierairegulationmanaging}. Administrative law suggests that delegating threshold setting could be a reasonable response to this complexity as long as sufficiently clear ("intelligible principle"-type) guidance is given to the AI and process requirements are followed. The ultimate decision to accept or reject the proposed threshold might still need to rest with an accountable human, but the technical details of research and implementation could be offloaded to some extent.

Second, AI companies must allocate limited red-teaming resources to evaluate the safety of their systems \cite{ahmad2025openaisapproachexternalred}. Determining how to deploy those resources is an optimization problem that requires synthesizing huge quantities of internal and external data on where risks might originate. This scenario might be one in which delegation to an AI is reasonable. AIs might be more competent at the relevant kinds of analysis and optimization than humans are, as well as better able to overcome their biases about where risks might be coming from. Administrative law-style checks like requiring explanations for resource allocation and human sign-off on huge shifts would help ensure that red-teaming stays on track and aligned with the overall goals of the company.

Third, AI companies are already creating mechanisms that determine how much  compute to allocate different users to solve their problems. Some allocation is simply done by pricing tiers, but consumer apps like ChatGPT now incorporate "routers" for allocation \cite{olaughlin2025gpt5}. Access to compute might someday become an essential part of modern life, so control over who can access it and under what circumstances will be highly significant. Administrative agencies must follow procedural rules when distributing essential benefits because of how important they are for people's lives \cite{reich1964new}. It may soon become necessary for AI companies or governments to similarly implement rules governing how compute is distributed to people and laying out whether and how users might have rights to compute that they can enforce against those seeking to limit or rescind access. Administrative law's doctrines could provide inspiration for ways to protect people in the novel contexts of AI, drawing on old lessons for these new problems.

\section{Limitations}
The foregoing analysis is preliminary and not meant to provide final rules for applying administrative law's lessons to AI governance. Both fields are deep and rapidly changing, as AI develops and the Supreme Court reconsiders foundational parts of administrative law doctrine. There are many areas where administrative law does not fit the kinds of problems that AI governance is facing. This paper is intended to illustrate the kinds of connections that implicitly exist between law and AI and start a conversation drawing on them to help improve each. Further work formalizing these insights through the creation of decision rules for delegation and the like would represent a helpful path forward.

\section{Conclusion}
Administrative law has its challenges and is currently undergoing a period of serious upheaval. Even at its best, it is an imperfect set of often vague rules trying to handle the massive challenges of administrative governance. Yet this imperfect body of doctrine has guided decades of controversial delegation. AI governance can learn key lessons of governance design from administrative law, and determining when and how to use AI in governance is one important place to start.

\newpage
\bibliographystyle{plain}
\bibliography{AI_Admin_Delegation}


\begin{enumerate}

\item {\bf Claims}
    \item[] Question: Do the main claims made in the abstract and introduction accurately reflect the paper's contributions and scope?
    \item[] Answer: \answerYes{}{} 
    \item[] Justification: The abstract and introduction clearly state the claims made, the framework of analysis, and the key contributions the paper is intended to make.

\item {\bf Limitations}
    \item[] Question: Does the paper discuss the limitations of the work performed by the authors?
    \item[] Answer: \answerYes{}{} 
    \item[] Justification: The paper specifically discusses the limitations of attempting to draw lessons from administrative law for AI, in the "Limitations" section and throughout.

\item {\bf Theory assumptions and proofs}
    \item[] Question: For each theoretical result, does the paper provide the full set of assumptions and a complete (and correct) proof?
    \item[] Answer: \answerNA{}{} 
    \item[] Justification: The paper does not include theoretical results.

    \item {\bf Experimental result reproducibility}
    \item[] Question: Does the paper fully disclose all the information needed to reproduce the main experimental results of the paper to the extent that it affects the main claims and/or conclusions of the paper (regardless of whether the code and data are provided or not)?
    \item[] Answer: \answerNA{} 
    \item[] Justification: The paper does not include experiments.
    
\item {\bf Open access to data and code}
    \item[] Question: Does the paper provide open access to the data and code, with sufficient instructions to faithfully reproduce the main experimental results, as described in supplemental material?
    \item[] Answer: \answerNA{} 
    \item[] Justification: The paper does not include experiments requiring code.

\item {\bf Experimental setting/details}
    \item[] Question: Does the paper specify all the training and test details (e.g., data splits, hyperparameters, how they were chosen, type of optimizer, etc.) necessary to understand the results?
    \item[] Answer: \answerNA{} 
    \item[] Justification: The paper does not include experiments.

\item {\bf Experiment statistical significance}
    \item[] Question: Does the paper report error bars suitably and correctly defined or other appropriate information about the statistical significance of the experiments?
    \item[] Answer: \answerNA{} 
    \item[] Justification: The paper does not include experiments.

\item {\bf Experiments compute resources}
    \item[] Question: For each experiment, does the paper provide sufficient information on the computer resources (type of compute workers, memory, time of execution) needed to reproduce the experiments?
    \item[] Answer: \answerNA{} 
    \item[] Justification: The paper does not include experiments.
    
\item {\bf Code of ethics}
    \item[] Question: Does the research conducted in the paper conform, in every respect, with the NeurIPS Code of Ethics \url{https://neurips.cc/public/EthicsGuidelines}?
    \item[] Answer: \answerYes{}{} 
    \item[] Justification: As a theoretical contribution from law, the paper does not implicate any ethical concerns.

\item {\bf Broader impacts}
    \item[] Question: Does the paper discuss both potential positive societal impacts and negative societal impacts of the work performed?
    \item[] Answer: \answerYes{}{} 
    \item[] Justification: The paper considers how the delegation of governance to AI systems might cause severe harms and undermine the extent of human participation in governance.
    
\item {\bf Safeguards}
    \item[] Question: Does the paper describe safeguards that have been put in place for responsible release of data or models that have a high risk for misuse (e.g., pretrained language models, image generators, or scraped datasets)?
    \item[] Answer: \answerNA{} 
    \item[] Justification: The paper poses no such risks.
    
\item {\bf Licenses for existing assets}
    \item[] Question: Are the creators or original owners of assets (e.g., code, data, models), used in the paper, properly credited and are the license and terms of use explicitly mentioned and properly respected?
    \item[] Answer: \answerNA{} 
    \item[] Justification: The paper does not use existing assets.

\item {\bf New assets}
    \item[] Question: Are new assets introduced in the paper well documented and is the documentation provided alongside the assets?
    \item[] Answer: \answerNA{} 
    \item[] Justification: The paper does not release new assets.

\item {\bf Crowdsourcing and research with human subjects}
    \item[] Question: For crowdsourcing experiments and research with human subjects, does the paper include the full text of instructions given to participants and screenshots, if applicable, as well as details about compensation (if any)? 
    \item[] Answer: \answerNA{} 
    \item[] Justification: The paper does not involve crowdsourcing nor research with human subjects.

\item {\bf Institutional review board (IRB) approvals or equivalent for research with human subjects}
    \item[] Question: Does the paper describe potential risks incurred by study participants, whether such risks were disclosed to the subjects, and whether Institutional Review Board (IRB) approvals (or an equivalent approval/review based on the requirements of your country or institution) were obtained?
    \item[] Answer: \answerNA{} 
    \item[] Justification: The paper does not involve crowdsourcing nor research with human subjects.

\item {\bf Declaration of LLM usage}
    \item[] Question: Does the paper describe the usage of LLMs if it is an important, original, or non-standard component of the core methods in this research? Note that if the LLM is used only for writing, editing, or formatting purposes and does not impact the core methodology, scientific rigorousness, or originality of the research, declaration is not required.
    \item[] Answer: \answerNA{} 
    \item[] Justification: The core method development in this research does not involve LLMs as any important, original, or non-standard components. We have only used LLMs for editing and formatting.

\end{enumerate}

\end{document}